\documentclass[twocolumn,a4paper,prb,showkeys,showpacs,superscriptaddress]{revtex4-1}
\usepackage{natbib}
\usepackage[final]{graphicx}
\usepackage{amsfonts}
\usepackage{amsmath}

\usepackage{amssymb}
\usepackage{color}
\usepackage{xcolor}
\usepackage{verbatim}
\usepackage{float}

\newcommand{\uam}{\affiliation{Dpto. Fisica Materia Condensada, C03, Universidad Autonoma de Madrid, 28049, Madrid, Spain}}
\newcommand{\dmit}{\affiliation{Francis Bitter Magnet Laboratory, Massachusetts Institute of Technology, Cambridge, Massachusetts 02139, USA}}
\newcommand{\dmituam}{\uam\dmit}
\newcommand{\dphys}{\affiliation{Department of Physics, National Taiwan University, Taipei 10617, Taiwan}}
\newcommand{\iams}{\affiliation{Institute of Atomic and Molecular Sciences, Academia Sinica, Taipei 10617, Taiwan}}
\newcommand{\tw}{\dphys\iams}
\newcommand{\dptu}{\affiliation{Department of Physics, Tamkang University, New Taipei City 25137, Taiwan}}
\newcommand{\tww}{\dphys\dptu}
\newcommand{\twww}{\dphys}
\newcommand{\dipc}{\affiliation{Donostia International Physics Center (DIPC), E-20018 Donostia-San Sebasti\'an, Spain}\affiliation{IKERBASQUE, Basque Foundation for Science, E-48013 Bilbao, Spain}}

\begin{document}

\title{Magnetic-state controlled molecular vibrational dynamics at buried molecular-metal interfaces}
\author{Isidoro Martinez$^\dagger$} \uam
\author{Juan Pedro Cascales$^\dagger$} \dmituam
\author{Cesar Gonzalez-Ruano} \uam
\author{Jhen-Yong Hong} \tww
\author{Chen-Feng Hung} \twww
\author{Minn-Tsong Lin} \tw
\author{Thomas Frederiksen}\dipc
\author{Farkhad G. Aliev} \email[Email: ]{farkhad.aliev@uam.es} \uam 

\date{\today}

\begin{abstract}

Self-assembled molecular structures have been intensively used in molecular electronics and spintronics. However, detailed nature of the interfaces between molecular layers and extended metallic contacts used to bias the real devices remains unclear. Buried interfaces greatly restrict application of standard techniques such as Raman or scanning electron microscopies. Here we introduce low frequency noise spectroscopy as a tool to characterize buried molecular-metal interfaces. We take advantage of vibrational heating of the molecules with incomplete contacts to the interface. Electrons, being the main spin and charge carriers propagating through the interfaces involving self-assembled molecules, interact inelastically with charged atomic ions. Such interactions produce quantum molecular vibrations (phonons). Detailed investigation of both conductance and conductance fluctuations in magnetic tunnel junctions with few nm Perylenetetracarboxylic dianhydride (PTCDA) allows to map vibrational heating at specific biases taking place in “hot spots” where self-assembled layers weaker contact the metallic electrodes. We follow this effect as a function of PTCDA thickness and find the best molecular-metal order for the lowest (3-5 monolayers) barriers. Moreover, we unveil interplay between spin and phonons at interface showing experimentally and by modelling spin-control over molecular vibrational heating. We find that vibrational heating related low frequency noise essentially depends on the relative alignment of the electrodes with noise changes well beyond expectations from fluctuation-dissipation theorem.

\end{abstract}

\keywords{Buried organic-metal interfaces; noise spectroscopy; molecular vibrational dynamics; spin control; modelling noise at organic-metal interface}

\maketitle

\section{Introduction}

Downscaling of inorganic electronics has currently become increasingly difficult because of physical limitations, due to surface-related modification of the electronic structures. The bottom-up concept introduced by molecular electronics in order to solve the problem has been around since the seventies \cite{Mann1971} and unique intrinsic features of molecules and quantum dots, among other excited phonons and vibrations \cite{Nitzan2003,Galperin2007,Kennehan2018}, have been explored intensively.

One of the major challenges in molecular electronics (molecular spintronics) is control over of molecules-metal (ferromagnetic electrodes) contacts \cite{Chen2018}. To avoid possible damage to molecular order and even pin-hole formation a deposition of seed under-oxidized atomic thick Alumina layer has been proposed \cite{Santos2007}. This method both improves interface quality, reduces influence of conductance mismatch and enhances spin life time \cite{Santos2007,Li2011}. However, as long as one deals with buried interfaces, the tools available to evaluate their quality are very limited. For such purpose (when electron transmission  microscopy is used \cite{Wiktor2017}) the device is usually destroyed.

Electron transport  through the interfaces  could be a non-destructive, albeit indirect method to evaluate buried molecule-metal interface quality. Few main factors determine transport in molecular electronics, with electron-phonon coupling being of importance along with molecular and electrodes electronic structures \cite{Galperin2005,Ward2011,Elliott2012}. The identification of phonon modes is usually carried out by Raman or inelastic tunneling spectroscopy (IETS) \cite{Elliott2012,Galbiati2015}. Due to strong photon attenuation by the metallic electrodes, Raman spectroscopy on heterostructures mainly reflects information from the out surface layers \cite{Tan2017}.  

The IETS technique, based on a weak change in the tunneling conductance beyond the vibrational onsets of excitation, is mostly used to explore molecular vibrations in the single-molecule limit\cite{Stipe1998,Smit2002} and when contacted by sharp tips with undefined magnetization direction.  For the organic magnetic tunnel junctions, the conductance signals are averaged over a large number of molecules forming either barrier or contacting extended ferromagnetic electrodes. To our best knowledge, the only study of the IETS in molecular spintronics just aimed to obtain structural information on the integrity of the $Alq_3$ barrier by localizing junctions with pin-holes where no signatures of molecular vibrational modes were present \cite{Galbiati2015}. As we shall see below, for the high quality buried molecular-metal interfaces with only small percentage on the molecules weakly contacting the electrode, the detailed investigation of the buried molecule-metal interfaces could be carried out by analyzing vibrational heating of the molecules in locally distributed “ hot spots”.

Vibrational heating (or cooling) of the molecules contacting metal electrodes is a well established phenomenon \cite{Dubi2011} for atomic and molecular junctions.  Vibrational heating at finite bias is among the major obstacles for advances in molecular electronics and spintronics due to the associated potential device breakdown \cite{Dubi2011,Simine2012,Galperin2007,Weiss2015}. Peculiarly, as mentioned above, a current flux can either heat or cool locally the molecular junction close to sub-vibration resonance conditions \cite{Galperin2009,Simine2012}.

The inclusion of the electron spin degree of freedom in organic magnetic tunnel junctions (OMTJs)\cite{Dediu2002,Xiong2004,Rocha2005,Santos2007,Sanvito2011,Jian2011,Vincent2012,Nguyen2012,Rakh2016} and interface-related molecular magnetism \cite{Sanvito2010,Lach2012} potentially allows spin control over vibrons, as pointed out in recent theoretical studies \cite{Droghetti2015,Lunghi2017}. Other theories go even further and predict phonon-dressed spins to exist with effective magnetic moments of $0.024~\mu_B$ for single-phonon quanta with the original system being non-magnetic \cite{Shin2018}. The interplay between spins and phonons in molecular spintronics remains however obscured.

As we shall see below, the conductance fluctuations or inelastic noise probing vibron excitations  could present an alternative tool capable of identifying high quality buried molecules-metal interfaces. At low temperatures and biases not yet exciting molecular vibrations or defect states, (e.g. typically below 10mV) voltage fluctuations are dominated by shot noise providing information on interfacial electron-electron correlations \cite{Cascales2014}. At higher biases (typically exceeding phonon energies situated above 15mV), the low-frequency noise (LFN) is known to provide unique capability of identifying surface located molecules through their vibrations in non-spintronic systems \cite{Tsutsui2010,Schaffert2012}.

Here we describe qualitative step further and introduce vibron noise spectroscopy, which addresses the challenge of investigation of quality of buried molecular-metal (ferromagnet) interfaces in OMTJs. Our experiments monitor conductance, tunneling magnetoresistance and voltage fluctuations as a function of applied bias and magnetic state in the OMTJs with efficient heat transport away from the molecules so that the junction's structure is not jeopardized. The junctions with thinnest (3-5 PTCDA monolayers) barrier show the smallest background noise and highest TMR of 20-30$\%$ indicative of their major structural order. In those junctions, we experimentally observe and theoretically model the spin-dependent molecular vibrational dynamics. The physical origin of this effect is related to the difference of available phase space for the inelastic tunneling process between parallel and antiparallel states. Besides, our results open the possibility of  spin heat-dissipation control in molecular spintronics.

\section{Results}

\begin{figure}
\begin{center}
\includegraphics[width=7cm]{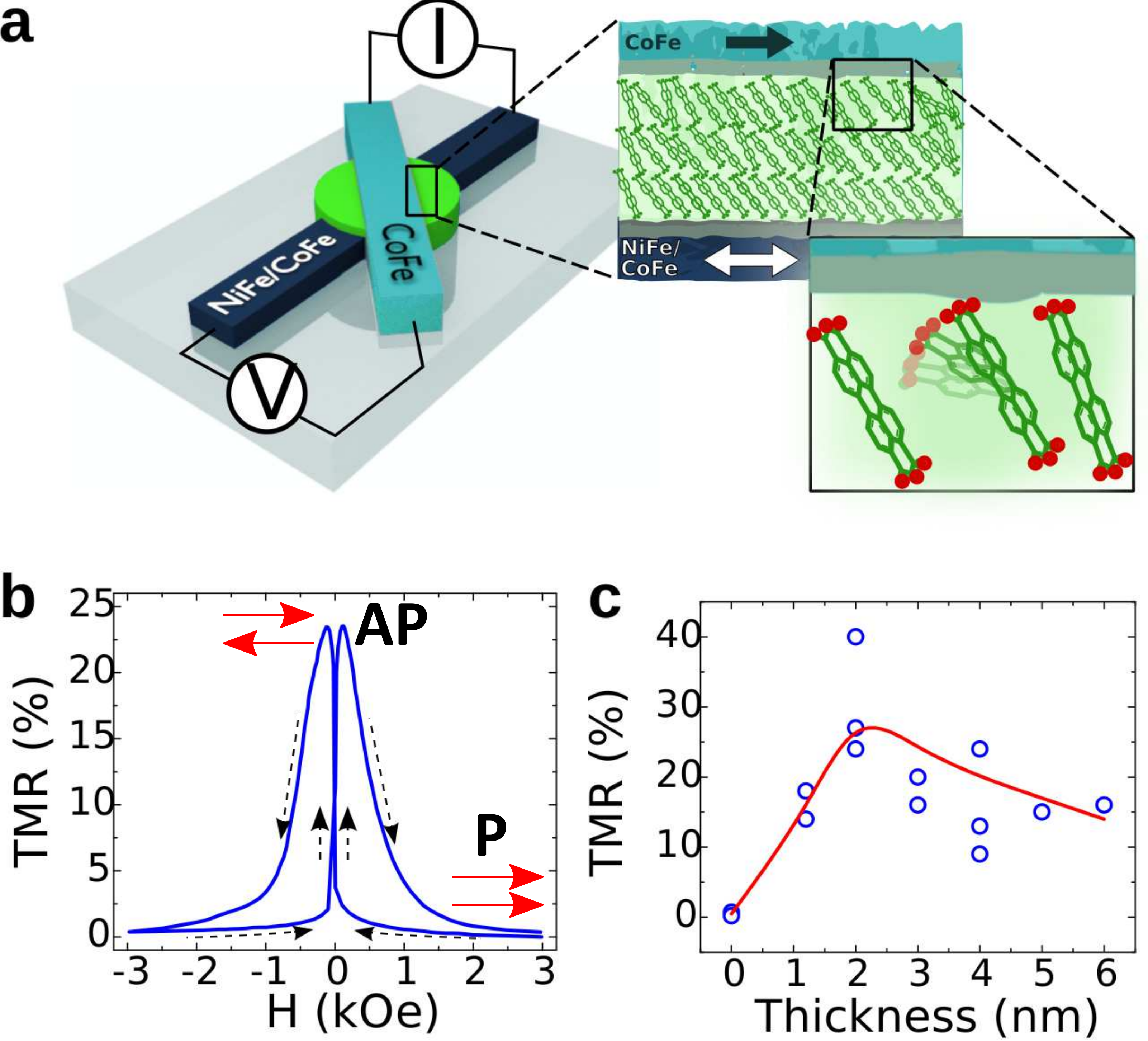} 
\end{center}
\caption{(a) Sketch of the junctions structure and zoom of the cross section of PTCDA barrier highlighting the possible molecular bond rattling due to interface roughness. (b) Typical low-bias (0.7 mV) TMR measured in OMTJs with a 2 nm thick barrier at 10 K. Red arrows indicate P and AP state while dotted black arrows show magnetic field sweep directions. (c) Dependence of the low-bias TMR ($T=10$ K) on the PTCDA thickness in OMTJs. The red curve is a guide for the eye.}
\label{fig:Fig1}
\end{figure}

The sample growth and measurement technique are explained in the Methods section. More details could be found in Refs \cite{Li2011,Hong2014,Herranz2010,Cascales2014}. 
Figure \ref{fig:Fig1} illustrates the investigated OMTJs highlighting the 1.2 nm thick barrier cross-section (with approximately three layer of PTCDA). The sketched molecular order (zoom in Fig.~\ref{fig:Fig1}a) describes a realistic PTCDA structure with some conductance channels at zero bias being open due to molecular bending, interface roughness or local disorder.

An exponential dependence of the room temperature conductance with the PTCDA thickness\cite{Cascales2014} and a large tunneling magnetoresistance (TMR) (Fig.~\ref{fig:Fig1}b,c) indicate that the electron transport through the PTCDA layer is dominated by direct tunneling with energy and spin relaxation close to the molecule/ferromagnet interfaces. Figure \ref{fig:Fig1}b shows two well-defined and reproducible magnetic states corresponding to parallel (P) and close to antiparallel (AP) alignment of the electrodes. Tunneling magnetoresistance values (up to $30\%$) observed with below 2nm thick PTCDA barriers (3-5 monolayers) are among the highest reported so far in molecular spintronic junctions in the tunneling regime. High TMR signals well ordered self-assembled thinnest PTCDA barrier. Conductance and noise in a total of 13 samples were investigated with PTCDA thickness ranging from 0 to 6 nm. Unless otherwise indicated, the measurements temperature is 10K.

\begin{figure}
\begin{center}
\includegraphics[width=1\linewidth]{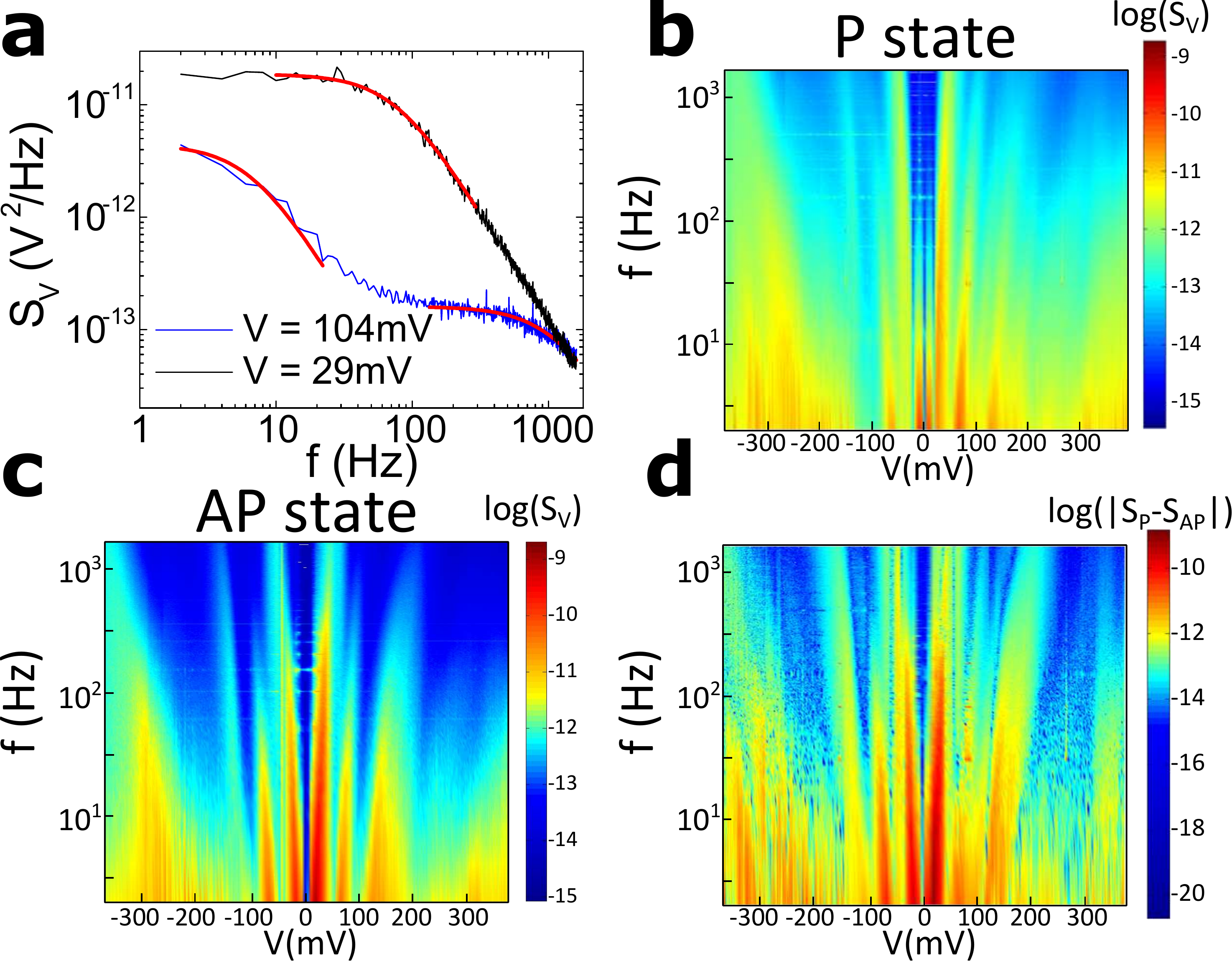}
\end{center}
\caption{Part(a) shows LFN spectra for two different biases close to the range with excess RTN with fit to a Lorentzian function (red curves). Parts (b) and (c) show the bias dependent LFN in the junction with 1.2 nm PTCDA at P and AP states respectively. Finally, the part (d) shows the absolute value of the difference of noise power in two magnetic states. The measurements have been carried out at $T=10$ K.}
\label{fig:Fig2}
\end{figure}

Figures \ref{fig:Fig2}a-c describe representative behavior of noise for OMTJs with PTCDA thickness of 1.2 nm in two magnetic states. These junctions show the lowest contribution to the $1/f$ noise signalling a higher molecular order. One observes unexpected enhancement of the noise power due to appearance of random telegraph noise (RTN) in conductance close to set of specific biases. Qualitatively similar magnetic state dependent LFN has been found when the measurements are carried out at base temperature of $T=0.3$ K (Fig.~S1, Ref.\cite{SupMaterial}). The close to symmetric character of the LFN anomalies with voltage $V$ points towards a \textit{vibrational origin} rather than to a resonant electron tunneling process, for which LFN enhancement is expected for one bias polarity only \cite{Clement2007}.

In order to analyze in detail the shape and symmetry of the RTN on bias, the observed effect has been represented in two qualitatively different plots. From the one side, Fig.~\ref{fig:Fig2}b shows rough data, \textit{i.e.}, the variation of the noise power in the parallel state ($S_\mathrm{P}$) with frequency and the applied bias. On the other hand, we have quantified how RTN characteristic frequency varies with applied bias and its polarity. As example, Fig.~\ref{fig:Fig2}a shows cross-sections of $S_\mathrm{P}(f,V)$ from the Figure \ref{fig:Fig2}b for the two particular biases close revealing specific RTN \textit{humps}. With increasing bias, the RTN first appears as a noise excess in the low-frequency limit which approximately varies as $1/(f^2 - f_{0}^2)$ with $f_0$ being the characteristic RTN frequency. This quantitative analysis confirms that set of specific RTN excitations is \textit{nearly symmetric in bias} and shows an increase of the $f_0$ under the applied bias before RTN disappears outside the experimental bandwidth. Some asymmetry could probably be related with differences in the bottom and top electrode roughness and electrode/barrier interface.

Interestingly, switching the device with clear LFN anomalies (here with PTCDA $\leq 2nm$ as we shall see below) to the AP state substantially changes the LFN (Fig.~\ref{fig:Fig2}c). In order to quantify the net change in LFN between the two magnetic states we calculate the absolute value of the difference of the noise power between P and AP states $|S_\mathrm{P}-S_\mathrm{AP}|$ (Fig.~\ref{fig:Fig2}d).
Note that for the both magnetic states and at biases exceeding 200 mV the RTN anomalies appear to be smeared out. As see in Fig.~\ref{fig:Fig2}b,c and specially for OMTJs with yet thicker PTCDA, in the high-bias range the so-called $1/f$ background noise becomes superimposed on the RTN contribution making it more challenging to accurately determine the RTN characteristic frequencies $f_0$

\begin{figure}
\begin{center}
\includegraphics[width=1\linewidth]{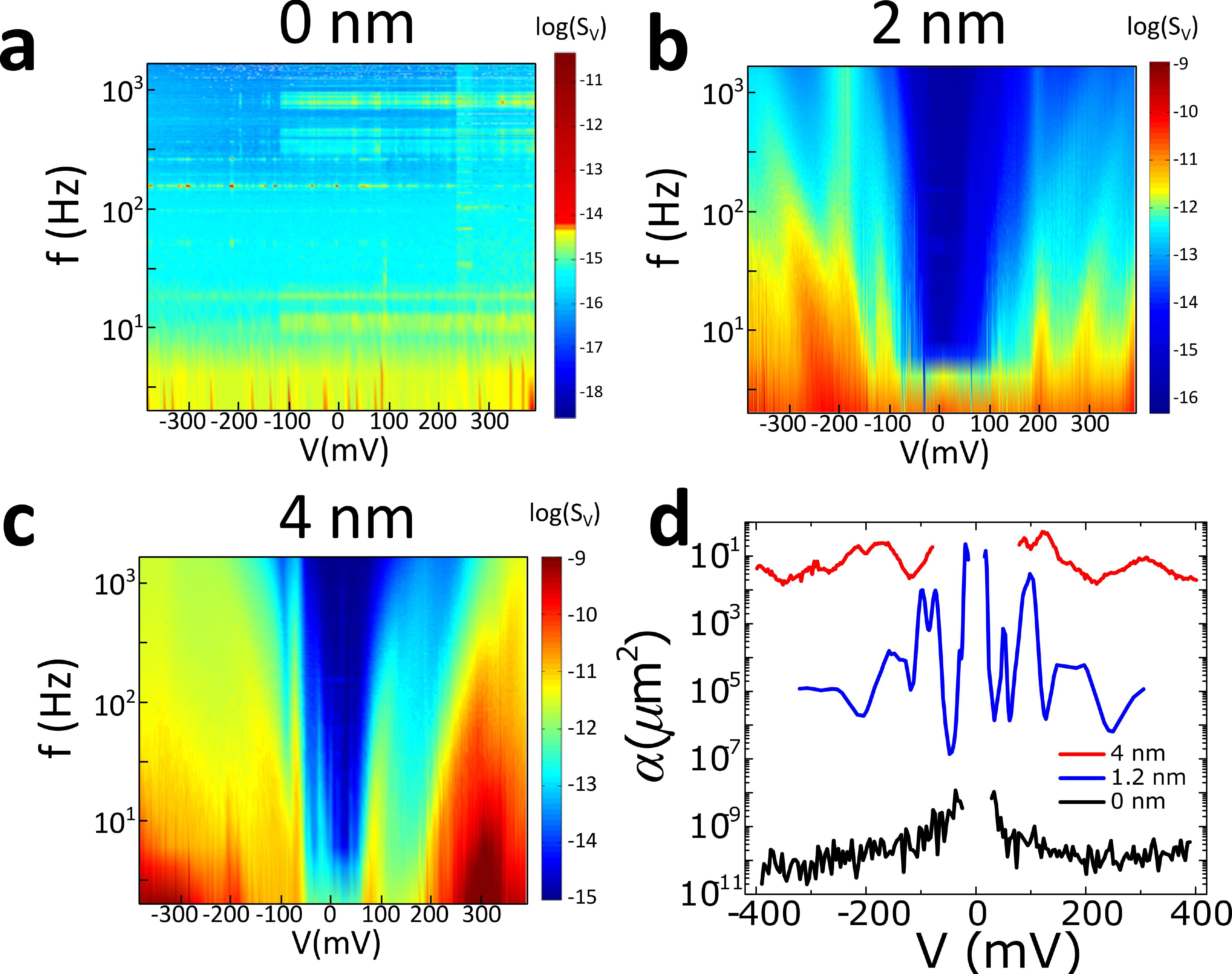}
\end{center}
\caption{Noise spectroscopy of OMTJs in the P state as a function of PTCDA thickness. Part(a) shows LFN spectra for control sample without PTCDA. Parts (b,c) show noise spectroscopy of OMTJ with 2 and 4nm PTCDA correspondingly. Finally, the part (d) shows variation of the bias dependent LFN in OMTJs at different PTCDA thicknesses measured at T=10 K and in the P state and quantified using the Hooge factor.}
\label{fig:Fig2a}
\end{figure}

In order to understand better the nature of the observed bias and magnetic state dependent RTN, we carried out noise spectroscopy as a function of PTCDA thickness ( see Fig.~\ref{fig:Fig2a}). Let us first analyse LFN in the control samples (\textit{i.e.}, junctions without PTCDA but with 0.6+0.6 nm of AlO$_x$ separating two ferromagnetic electrodes). One observes practically featureless background noise (Fig.~\ref{fig:Fig2a}a) which is at least two orders in the magnitude below the LFNs measured in any of the junctions with a PTCDA barrier. We conclude that set of the bias-dependent RTN features originates from molecular vibrations inside the PTCDA barrier or at the PTCDA/electrode interfaces.

Further increase of the PTCDA thickness maintains the well-defined bias dependent RTN anomalies only for barrier thickness below or about 2nm. We observe the clear signatures of vibron related LFN with 1.2nm and partially with 2nm PTCDA thickness only. The further increase of the PTCDA thickness (as shown for OMTJ with 4nm PTCDA) smeares out vibron related signatures in noise due to onset of the strong background 1/f noise already at relatively small biases (Fig.~\ref{fig:Fig2a}c).

To show more clear the applicability range of the vibron noise spectroscopy when following the qualitative changes in the LFN with increasing PTCDA thickness we may also use the Hooge factor explained in Methods. Fig.~\ref{fig:Fig2a}d plots the bias dependence of the Hooge factor for OMTJs with three different PTCDA thicknesses. While the control samples show a practically featureless Hooge vs.~bias dependence, the introduction of 1.2 nm PTCDA (about three PTCDA layers) between two ferromagnetic electrodes raises dramatically the normalized noise (several orders of magnitude). The strongest relative LFN anomalies with bias were observed for the 1.2-2 nm thick PTCDA barriers (Fig.~\ref{fig:Fig2a}b,d). A further increase of the PTCDA thickness boosts the LFN background, approaching the LFN spectrum closer to the $1/f$ type. This induces a relative decrease of the magnitude of LFN anomalies, as seen for the OMTJs with 4 nm thick PTCDA. Generally, the evolution of LFN with PTCDA points towards a predominantly interfacial origin of the LFN anomalies, gradually affected by the accumulation of molecular disorder with an increasing barrier thickness.

It is important to note that spin-dependent molecular vibron noise (Fig.~\ref{fig:Fig2}) in few orders of the magnitude exceeds in the noise power expected from the application of fluctuation-dissipation theorem (FDT). Indeed, from FDT the thermal noise power is \cite{Alievbook2018} $S_V(f)=4k_BTR$ (with $T$ being temperature, $R$- junction resistance and $k_B$- Boltzmann contant).  Then below 10K with measured resistance change between P and AP state of few kOhm, the noise power change expected from FDT should be less than $10^{-18}\; \mathrm{V}^2/\mathrm{Hz}$.  The experimentally observed variation of the noise power between P and AP states (see Figs.~\ref{fig:Fig2}d) exceeds that by more than six orders of the magnitude.

Concerning the robustness of the bias-dependent RTN anomalies in each magnetic state we have carried out a number of reproducibility tests. The corresponding analysis shown in Fig.~S2 (Ref.\cite{SupMaterial}) permits to conclude that the RTN vs.~bias anomalies are rather reproducible.

Interestingly, the enhancement of the LFN power at specific biases (up to several orders of the magnitude) is accomplished with limited impact (with relative change below $10^{-2}$) on the bias-dependent conductance (Fig.~S3, Ref.\cite{SupMaterial}). This suggests that a great majority of the molecular conductance channels in our OMTJs arise from well-established contacts between molecules and the electrodes. At the same time, only a small fraction of the total conductance channels (roughly estimated to be $10^{-2} - 10^{-3}$ from the relation between square root of the integrated over experimental bandwidth voltage noise power to the applied bias) contribute to the RTN as schematically sketched in Fig.~\ref{fig:Fig1}a.

\begin{figure}
\begin{center}
\includegraphics[width=0.8\linewidth]{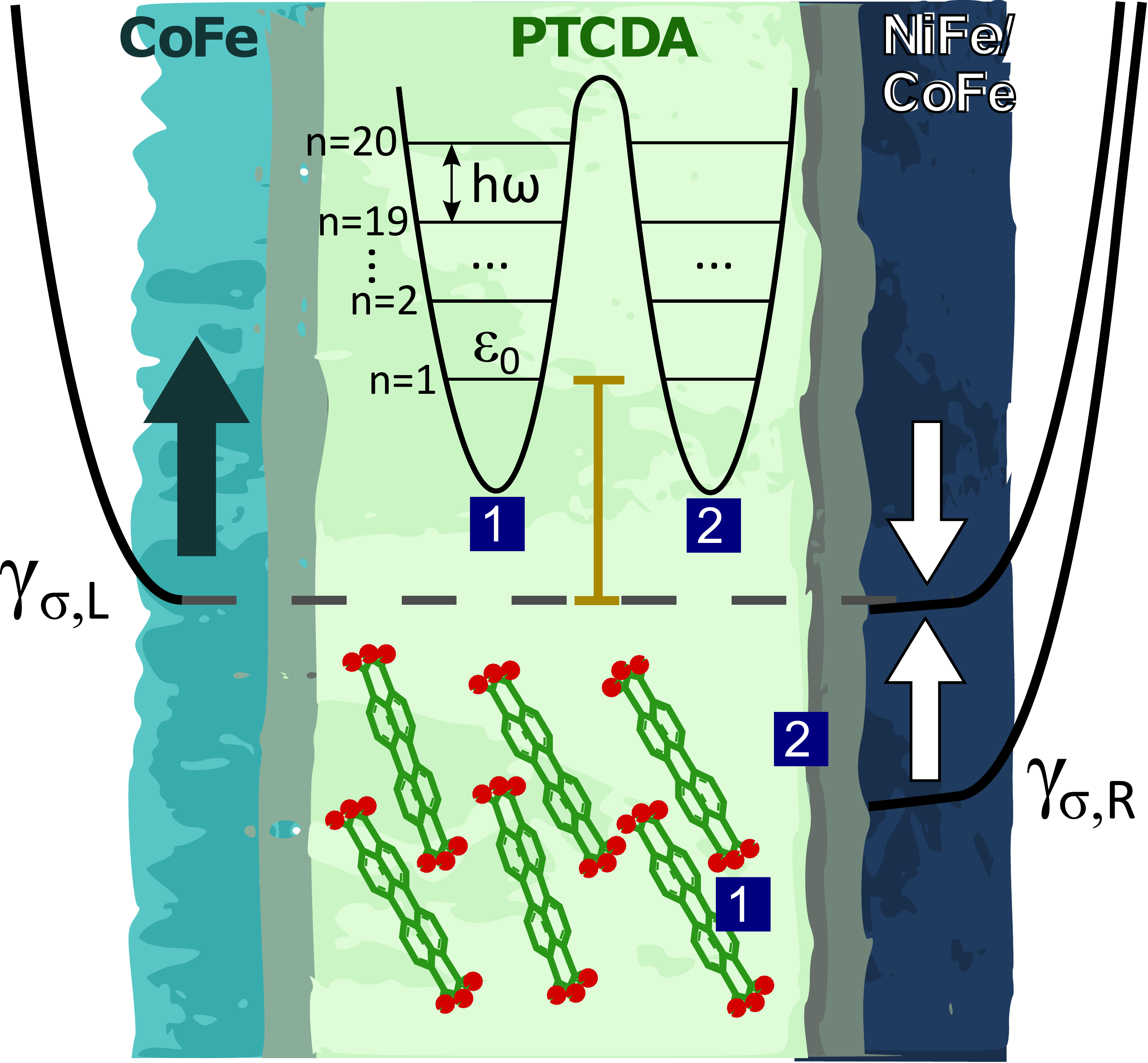}
\end{center}
\caption{Model for two state switch driven by inelastic excitations by spin polarised current and parameters used. We consider symmetric junction in the off-resonant regime with one-electron energy $\epsilon_0=10\gamma_L=10\gamma_R=1$eV and $\gamma_{L,R}$ being spin dependent density of states for left (L) and right (R) electrodes correspondingly. We used reaction order $n=20$ and coupling strength $\chi=1$ meV. The electrode spin polarization ($\sigma$) is depicted by the vertical black arrows.}
\label{fig:Fig3}
\end{figure}

\section{Discussion}

Based on this scenario, we consider below a simple model of the barrier in which a molecular conduction pathway can attain two different states, e.g., due to unavoidable interface roughness (Fig.~\ref{fig:Fig3}). The configuration becomes unstable under applied bias and switching between, for example, straight and bended molecular conformal configurations may occur. 
The stable conduction pathways (in parallel) contribute only to the background signal and provide practically a featureless $1/f$ noise. We therefore suggest that the observed excess of the RTN above certain characteristic biases is a result of inelastic excitations and conformational switching of individual molecules (or collective groups) in the barrier.

Following Refs.~\onlinecite{Gao1997, Brumme2012, Frederiksen2014} we generalize a model for a two-state switch driven by inelastic excitations by a \emph{spin-polarized} electronic current with main parameters described in Fig.~\ref{fig:Fig3}, the main text and detailed in SM \cite{SupMaterial}. Interestingly, this model predicts that the reaction yield, \textit{i.e.}, the probability that a tunneling electron induces the reaction, is \emph{larger} in the P than in the AP electrode configuration (Fig.~S6). The origin of this effect is related to the available phase space (initial and final states) for the inelastic tunneling process which is larger in P than in AP. This is analogous to the TMR effect where the elastic transmission of electrons is also larger in P than AP. We note that these two effects compound in the reaction rate $R$ (Eq.~S19, Ref.\cite{SupMaterial}).

\begin{figure}
\begin{center}
\includegraphics[width=1\linewidth]{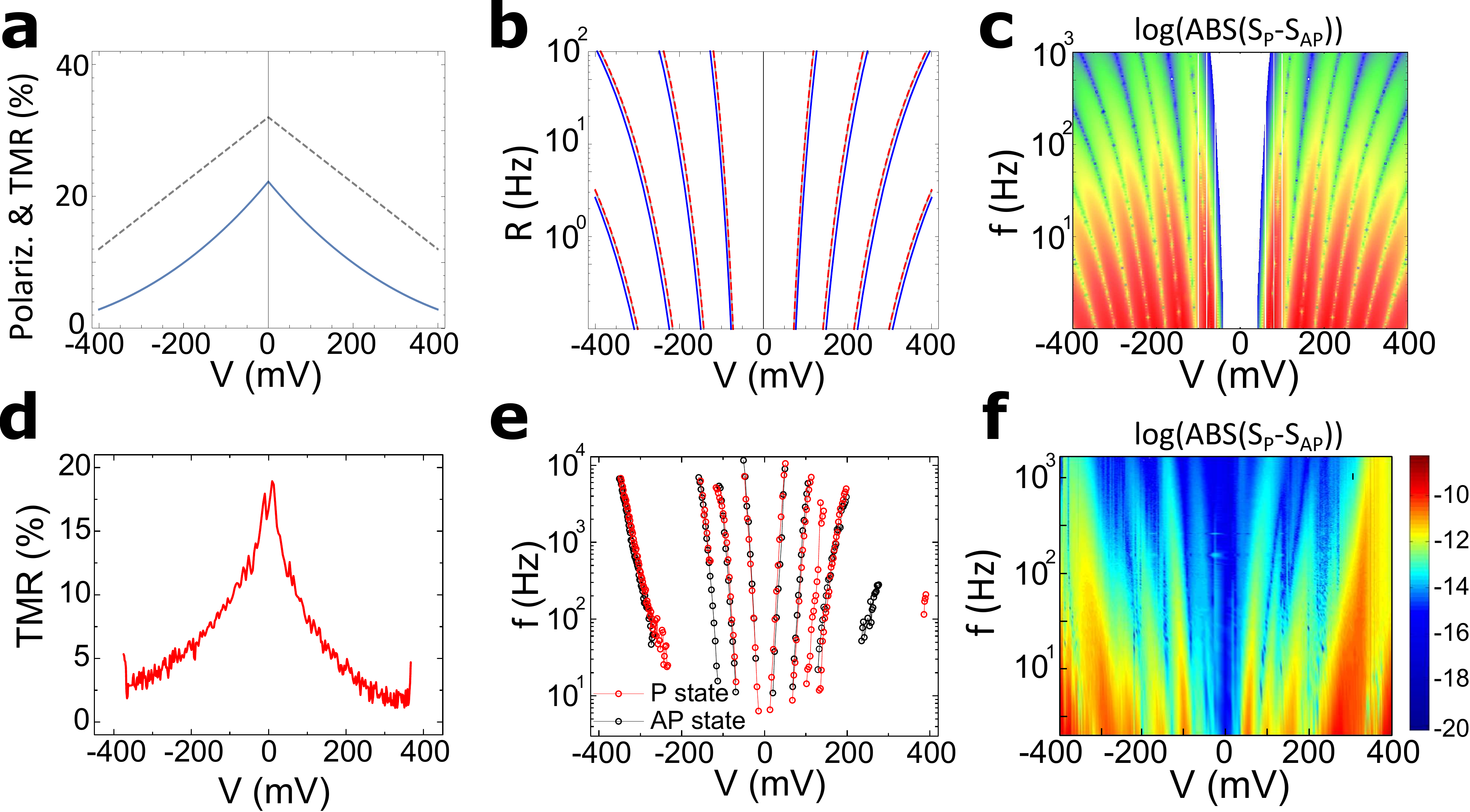}
\end{center}
\caption{Comparison between LFN modeling and experiment for P and AP states. a) Model for the electrode polarizations $\mathcal{P}_\alpha(V)$ (dotted grey line, Eq.~S16 Ref. \cite{SupMaterial}) and TMR (solid blue line, Eq.~S22, Ref.\cite{SupMaterial}) vs.~applied bias. b) Calculated switching rates $R$ in both P (dotted red line), AP (full blue line), and unpolarized (full grey line) cases. c) Power noise spectrum difference between P and AP states in absolute value. Measurements of d) TMR at $T=10$ K, e) RTN characteristic frequencies in P (red dots) and AP (black dots) states at $T=10$ K and f) power noise spectrum at $T=0.3$ K. Measurements were made in the 1.2 nm thick PTCDA sample.}
\label{fig:Fig4}
\end{figure}

To link these findings to the experimental observations we present in Fig.~\ref{fig:Fig4} an application of the model with appropriately chosen parameters. As detailed in the SM \cite{SupMaterial} we consider the metal-molecule-metal tunnel junction as described by a single electronic resonance bridging two metal electrodes with coupling of the electrons to a set of vibrational modes on the molecular bridge. The numerical results shown in Fig.~\ref{fig:Fig4} were obtained with the
following model parameters defined in the SI: $\varepsilon_0=10\gamma_L=10\gamma_R=1$ eV, $n=20$ and $\chi=1$ meV. The molecular vibrations were described by a set of modes characterized by energies being multiples of $\hbar\Omega=20$ meV. The electrode polarization was defined phenomenologically as $\mathcal{P}_L=\pm\mathcal{P}_R=0.32-0.5 \mathrm{V}^{-1}|V|$. The resulting TMR and LFN spectrum $S(f)$ were computed using Eqs.~S22 and S26, respectively, with the switching rate $R$ given by Eq.~S19 (Ref. \cite{SupMaterial}).

Figure \ref{fig:Fig4}a shows the electrode polarization (dashed gray curve) and TMR (blue curve), where the latter describes well the typical experimental voltage dependence of the TMR (Fig.~\ref{fig:Fig4}d).
For simplicity we represent the molecular vibrations by a set of modes characterized by energies being multiples of $\hbar\Omega=20$ meV. At a given applied voltage this translates into a set of characteristic switching rates as shown in Fig.~\ref{fig:Fig4}b, qualitatively similar to the experimental data in Fig.~\ref{fig:Fig4}e. Finally, we analyze the magnetic-state dependent noise spectrum $S=S(V,f)$ for realistic bias-dependent TMRs and associated spin polarizations. By the eye it is hard to appreciate the difference by P or AP (Fig.~\ref{fig:Fig4}b,e and Fig.~S4, Ref. \cite{SupMaterial}). The difference only becomes apparent when one considers the quantity $|S_\mathrm{P}-S_\mathrm{AP}|$ as shown in \ref{fig:Fig4}c, in some resemblance with the experimental data in \ref{fig:Fig4}f. We thus deduce that our modeling qualitatively can reproduce the main experimental observations.

In \textit{summary}, we have introduced low frequency noise spectroscopy as
a tool to characterize buried organic-metal interfaces. Key sensors for this spectroscopy are "hot spots" where molecular vibrational heating induces random telegraph noise detectable by standard noise measurement techniques. Moreover, we experimentally demonstrate and describe by the simple model the spin control of vibrational dynamics and switching in molecular spintronic devices. Our findings suggest the main LFN mechanism in molecular tunnel junctions with relatively ordered thin barriers and buried interfaces and therefore important for the development of reliable molecular spintronic devices.

\section{Methods}

\subsection{Sample growth}

The layer sequence in the PTCDA organic spin valves studied in this work is: NiFe(25 nm)\slash CoFe(15 nm)\slash Al$_{x}$O(0.6 nm)\slash PTCDA(1.2-5 nm)\slash Al$_{x}$O(0.6 nm)/CoFe(30 nm). The structure was deposited onto a glass substrate, and the junctions were prepared in a high-vacuum environment with a base
pressure lower than $10^{-8}$ mbar. The metallic layers were deposited by sputtering with an Ar working pressure of $5\times 10^{-3}$ mbar. The PTCDA layers were grown by thermal evaporation at $10^{-8}$ mbar, with a deposition rate of 0.1 nm/s. Thin AlO$_{x}$ layers were grown between the PTCDA layer and both ferromagnetic layers by partially oxidizing Al in an oxygen plasma for 5 s.

\subsection{Electron transport and noise measurements}

Electron transport and noise were measured down to 0.3 K using a He-3 JANIS cryostat. LFN experiments were carried out using a Stanford Research SR785 spectrum analyzer. The signal is duplicated into two channels identically amplified in two stages in order to use a cross-correlation technique so we may neglect extrinsic voltage noise contributions. We analyze the variation of LFN by direct plots of the noise power. For the qualitative comparison of the frequency-averaged LFN levels in our OMTJs with PTCDA barrier thickness we also used the Hooge factor $\alpha$ defined\cite{Herranz2010} via $S_V(f)=\alpha V^2/A f^{\beta}$ where $V$ is the bias voltage, $A$ the junction area, $f$ the frequency, and $\beta$ a parameter close to 1.

\section{Acknowledgements}

We thank Juan Carlos Cuevas for stimulating discussions and constructive comments. The work in Madrid has been supported by Spanish MINECO (MAT2015-66000-P and EUIN2017-87474), Comunidad de Madrid (NANOFRONTMAG-CM S2013/MIT-2850) and IFIMAC (MDM-2014-0377). TF acknowledges FIS2017-83780-P from the Spanish MINECO. We also gratefully acknowledge support by UAM-Santander collaborative project (2015/ASIA/04). J.P.C. acknowledges support from the Fundacion Seneca (Region de Murcia) postdoctoral fellowship (19791/PD/15) \vspace{5mm}

$^\dagger$ These authors contributed equally

\end{document}